\documentclass{elsart}

\usepackage{epsfig}
\usepackage{amssymb}

\begin{document}

\begin{frontmatter}

\title{Anisotropic stars with non-static conformal symmetry}

\author[label1]{Dibyendu Shee}\ead{dibyendu\_shee@yahoo.com},
\author[label2]{Farook Rahaman}\ead{rahaman@iucaa.ernet.in},
\author[label3]{B.K. Guha}\ead{bkguhaphys@gmail.com},
\author[label4]{Saibal Ray}\ead{saibal@iucaa.ernet.in}

\address[label1]{Department of Physics, Bengal Engineering
College Model School, B. Garden, Howrah 711103,West Bengal, India}
\address[label2]{Department of Mathematics, Jadavpur University, Kolkata
700032, West Bengal, India}
\address[label3]{Department of Physics,
Indian Institute of Engineering Science \& Technology, B. Garden,
Howrah 711103, West Bengal, India}
\address[label4]{Department of Physics, Government College of Engineering \& Ceramic
Technology, Kolkata 700010, West Bengal, India}

\date{today}

\begin{abstract}
We have proposed a model for relativistic compact star with anisotropy
and analytically obtained exact spherically symmetric solutions describing
the interior of the dense star admitting non-static conformal symmetry.
Several features of the solutions including drawbacks of the model have been
explored and discussed. For this purpose we have provided the energy conditions,
TOV-equations and other physical requirements and thus thoroughly investigated stability,
mass-radius relation and surface redshift of the model. It is observed that
most of the features are well matched with the compact stars, like quark/strange stars.
\end{abstract}

\begin{keyword}
General Relativity; noncommutative geometry; compact stars
\end{keyword}

\end{frontmatter}

\section{Introduction}

Since the striking idea of {\it white dwarf} by Chandrasekhar
\cite{Chandrasekhar1931} the study of general relativistic compact
objects received a tremendous thrust to carry out research in the
field of ultra-dense objects. White dwarfs are composed of one of
the densest forms of matter known, surpassed only by other compact
stars such as neutron stars, quark stars/strange stars, boson
stars, gravastars etc.

In the compact stars the matter is found to be in stable ground
state where the quarks are confined inside the hadrons. If it is
composed of the de-confined quarks then also a stable ground state
of matter, known as `strange matter', is achievable which provides
a `strange star'
\cite{Witten1984,Glendenning1997,Xu2003,DePaolis2007}. The main
theoretical motivation for postulating the existence of strange
stars was to explain the exotic phenomena of gamma ray bursts and
soft gamma ray repeaters \cite{Cheng1996,Cheng1998}. On the
observational side, the Rossi X-ray Timing Explorer has now
confirmed that $SAX J1808.4-3658$ is one of the candidates for a
strange star \cite{Li1999}.

As a start of the astrophysical objects of compact nature of the
above kind the theoretical investigation has been done by several
researchers by using both analytical and numerical methods.
However, the density distribution inside the compact stars need
not be isotropic and homogeneous as proposed in the TOV model. In
the early seventies Ruderman \cite{Ruderman1972} has investigated
the stellar models and argued that the nuclear matter may have
anisotropic features at least in certain very high density ranges
($> ~ 10^{15}$gm/c.c.), where the nuclear interaction must be
treated relativistically. Later on Bowers and Liang
\cite{Bowers1974} gave emphasis on the importance of locally
anisotropic equations of state for relativistic fluid spheres.
They showed that anisotropy might have non-negligible effects on
such parameters like maximum equilibrium mass and surface
redshift. Anisotropic matter distribution have been considered
recently by several investigators as key feature in the
configuration of compact stars (with or without cosmological
constant) to model the objects physically more realistic
~\cite{Mak2002,Mak2003,Usov2004,Varela2010,Rahaman2010a,Rahaman2011,Rahaman2012a,Kalam2012}.

As one of the physical characteristics Mak and Harko
\cite{MaK2000} have calculated the mass-radius ratio or
compactness factor for compact relativistic star however in the
presence of cosmological constant. To study mass and radii of
neutron star Egeland \cite{Egeland2007} incorporated the existence
of cosmological constant proportionality depending on the density
of vacuum. Kalam et al. \cite{Kalam2012} by using their model
calculated mass-radius ratio for Strange Star $Her~X-1$ of their
model and found it as ${M_{eff}/R}_{max} = 0.336$ which satisfies
Buchdahl's limit and corresponds to the same mass-radius ratio for
the observed Strange Star $Her~X-1$. This physical aspect is very
important because of the fact it helps to determine nature of the
compact stars, its size, shape, matter contain and several other
features.

Now to get more information, several studies have been done on
charged or neutral fluid spheres with a spacetime geometry that
admits a conformal symmetry, in the static as well as generalized
non-static cases. In this line of conformal symmetry we note that
there are lots of works available in the literature
~\cite{Dev2004,Yavuz2005,Pradhan2007,Sharma2007,Rahaman2010b}.
However, we investigate in the present work a new anisotropic star
admitting non-static conformal symmetry (mathematical formalism is
provided in detail in the next Sec. 2).

Under this background and motivation we investigation in the
present paper a model for relativistic compact star with
anisotropy and find out exact spherically symmetric solutions
which describe the interior of the dense star admitting non-static
conformal symmetry. The scheme of this study is as follows: We
provide mathematical formalism for non-static conformal motion in
Sec. 2 whereas the Einstein field equations for anisotropic
stellar source are given in Sec. 3. The solutions and general
features of the field equations under the Dev-Gleiser
energy-density profile have been found out in Sec. 4 along with a
special discussion on two prescriptions, in section 5, viz. ({\it
5.1}) the Misner prescription, and ({\it 5.2}) the Dev-Gleiser
prescription are done to show that mass to size ratio clearly
indicates that the model represents a quark/strange compact star.
In Sec. 6 we have made some concluding remarks.

\section{NON-STATIC CONFORMAL SYMMETRY}
 Now, we consider the interior of a star under conformal motion
through non-static Conformal Killing Vector as
\cite{Maartens1990,Coley1994,Harko2005,Boehmer2007,Rahaman2010b,Radinschi2010}
\begin{equation}
L_{\xi}g_{ij}=g_{ij;k}\xi^{k}+g_{kj}\xi_{;i}^{k}+g_{ik}\xi_{;j}^{k}=\psi
g_{ij}, \label{eq1}
\end{equation}
where $L$ represents the Lie derivative operator, $\xi$ is the
four vector along which the derivative is taken, $\psi$ is the
conformal factor and $g_{ij}$ are the metric potentials
\cite{Maartens1990}. It is to be noted that for $\psi=0$, the
equation above yields the killing vector, whereas for
$\psi=const.$ corresponds to the homothetic vector. So, in general
for $\psi=\psi(x,t)$ we obtain conformal vectors. In this manner,
one can perform a more general study of the spacetime geometry by
using Conformal Killing Vector.

The static spherically symmetric spacetime is given by the line
element (with $G=1=c$ in geometrized units)
\cite{Ray2008,Usmani2011}

\begin{equation}
ds^{2}=-e^{\nu(r)}dt^{2}+e^{\lambda(r)}dr^{2}+r^{2}(d\theta^{2}+sin^{2}\theta
d\phi^{2}), \label{eq2}
\end{equation}
where $\nu(r)$ and $\lambda(r)$ are the metric potentials and
functions of radial coordinate $r$ only.

The proposed charged fluid spacetime is mapped conformally onto
itself along the direction $\xi$. Here following Herrera et al.
\cite{Herrera1984,Herrera1985} we assume $\xi$ as
non-static but $\psi$ to be static as follows:
\begin{equation}
\xi=\alpha(t,r)\partial_{t}+\beta(t,r)\partial_{r}, \label{eq3}
\end{equation}

\begin{equation}
\psi=\psi(r). \label{eq4}
\end{equation}

The above set of equations (1)-(4) give the following expressions
\cite{Maartens1990,Coley1994,Harko2005,Boehmer2007,Rahaman2010a,Radinschi2010}
for $\alpha$, $\beta$, $\psi$ and $\lambda$:

\begin{equation}
\alpha=A+\frac{1}{2}kt, \label{eq5}
\end{equation}

\begin{equation}
\beta=\frac{1}{2}Bre^{-\frac{\lambda}{2}},\label{eq6}
\end{equation}

\begin{equation}
\psi=Be^{-\frac{\lambda}{2}}, \label{eq7}
\end{equation}

\begin{equation}
e^{\nu}=C^{2}r^{2}exp\left[-2kB^{-1}\int\frac{e^{\frac{\lambda}{2}}}{r}dr\right],
\label{eq8}
\end{equation}
where $k$, $A$, $B$ and $C$ are arbitrary constants. According to
Maartens and Maharaj \cite{Maartens1990} one can set $A=0$ and
$B=1$ so that by rescalling we can get
\begin{equation}
\alpha=\frac{1}{2}kt, \label{eq9}
\end{equation}

\begin{equation}
\beta=\frac{1}{2}re^{-\frac{\lambda}{2}}, \label{eq10}
\end{equation}

\begin{equation}
\psi=e^{-\frac{\lambda}{2}}, \label{eq11}
\end{equation}

\begin{equation}
e^{\nu}=C^{2}r^{2}exp\left[-2k\int\frac{e^{\frac{\lambda}{2}}}{r}dr\right].
\label{eq12}
\end{equation}

However, one can note that the above considerations, i.e. $A=0$
and $B=1$, do not loss any generality. This is because the
rescalling $\xi$ and $\psi$ in the following manner,
$\xi\rightarrow B^{-1}\xi$ and $\psi\rightarrow B^{-1}\psi$,
leaves Eq. (1) invariant.

\section{THE FIELD EQUATIONS FOR ANISOTROPIC STELLAR SOURCE}
In the present investigation we consider the most general
energy-momentum tensor compatible with spherically symmetry in the
following form:
\begin{equation}
T_{\nu}^{\mu}=(\rho+p_{r})u^{\mu}u_{\nu}+p_{r}g_{\nu}^{\mu}+(p_{t}-p_{r})\eta^{\mu}\eta_{\nu},
\label{eq13}
\end{equation}
with $u^{\mu}u_{\mu}=-\eta^{\mu}\eta_{\mu}=1$, where $\rho$ is the
matter density, $p_{r}$ and $p_{t}$ are radial and the transverse
pressures of the fluid respectively.

For the metric given in Eq. (\ref{eq2}), the Einstein field
equations are \cite{Rahaman2010c}
\begin{equation}
e^{-\lambda}\left[\frac{\lambda^{\prime}}{r}-\frac{1}{r^{2}}\right]+\frac{1}{r^{2}}=8\pi\rho,
\label{eq14}
\end{equation}

\begin{equation}
e^{-\lambda}\left[\frac{\nu^{\prime}}{r} + \frac{1}{r^{2}}
\right]-\frac{1}{r^{2}}=8\pi p_{r}, \label{eq15}
\end{equation}

\begin{equation}
\frac{1}{2}e^{-\lambda}\left[\frac{1}{2}(\nu^{\prime})^{2}+\nu^{\prime\prime}-\frac{1}{2}\lambda^
{\prime}\nu^{\prime}+\frac{1}{r}(\nu^{\prime}-\lambda^{\prime})\right]=8\pi
p_{t}. \label{eq16}
\end{equation}

Imposing the conformal motion, one can write the stress energy
components in terms of conformal function as follows:

\begin{equation}
8\pi\rho=\frac{1}{r^{2}}(1-\psi^{2}-2r\psi{\psi^{\prime}}),
\label{eq17}
\end{equation}

\begin{equation}
8\pi p_{r}=\frac{1}{r^{2}}(3\psi^{2}-2k\psi-1), \label{eq18}
\end{equation}

\begin{equation}
8\pi p_{t}= \frac{1}{r^{2}}\left[(\psi - k)^2
+2r\psi{\psi^{\prime}}\right]. \label{eq19}
\end{equation}

\section{THE SOLUTIONS AND FEATURES OF THE FIELD EQUATIONS
FOR DEV-GLEISER ENERGY-DENSITY PROFILE}

Now our task is to find out solutions for the above set of the
modified Einstein equation in terms of the conformal motion.
However, to do so we would like to study the following case
for specific energy density as proposed by Dev and Gleiser \cite{Dev2004}
for stars with the energy density in the form
\begin{equation}
\rho= \left(\frac{a}{r^{2}}+3b\right)/8\pi,\label{eq-rho}
\end{equation}
where $a$ and $b$ are constants. It is to be noted that the choice
of the values for $a$ and $b$ is dictated by the physical
configuration under consideration, e.g. $a = 3/7$ and $b = 0$,
corresponds to a relativistic Fermi gas for ultradense cores of
neutron stars \cite{Misner1964} whereas the values $a = 3/7$ and
$b \neq 0$ represent a relativistic Fermi gas with core immersed
in a constant density background \cite{Dev2004}. It can easily be
observed from Eq. (\ref{eq-rho}) that for large $r$ the constant
density term dominates ($r^2_c \gg a/3b$) so that the
corresponding star can be modeled with shell-type feature
surrounding the core.

Due to use of the above energy density, Eq. (\ref{eq17}) will take
the following form:
\begin{equation}
\left(\frac{a}{r^{2}}+3b\right)=
\frac{1}{r^{2}}\left(1-\psi^{2}-2r\psi{\psi{\prime}}\right),
\label{eq20}
\end{equation}

Solving this Eq. (\ref{eq20}), we get
\begin{equation}
\psi^{2}=1-a-b{r^2}+\frac{d}{r}= g(r)~(say), \label{eq21}
\end{equation}
where $d$ is an integration constant.

Substituting the value of $\psi$ in Eqs. (\ref{eq18}) and (\ref{eq19}),
expressions for radial and tangential pressures can be obtained as
\begin{equation}
p_{r}= \frac{1}{8\pi{r^2}}\left[2-3a-3br^2+\frac{3d}{r} -2k\sqrt
{g(r)}\right], \label{eq22}
\end{equation}

\begin{equation}
p_{t}= \frac{1}{8\pi{r^2}}\left[1-a-3br^2+k^{2}- 2k\sqrt
{g(r)}\right]. \label{eq23}
\end{equation}

\begin{figure}[htbp]
\includegraphics[scale=.3]{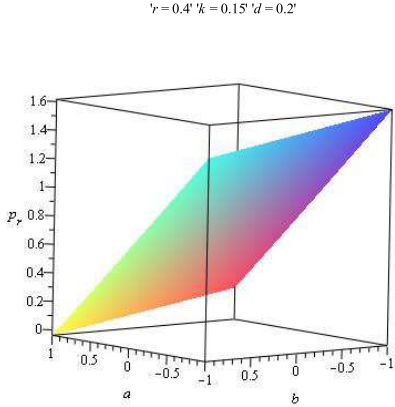}
\caption{Variation of radial pressure, $p_r$, with respect to the
parameters $a$ and $b$ for a fixed radial distance  $r=0.4$ km}
\end{figure}

\begin{figure}[htbp]
\includegraphics[scale=.3]{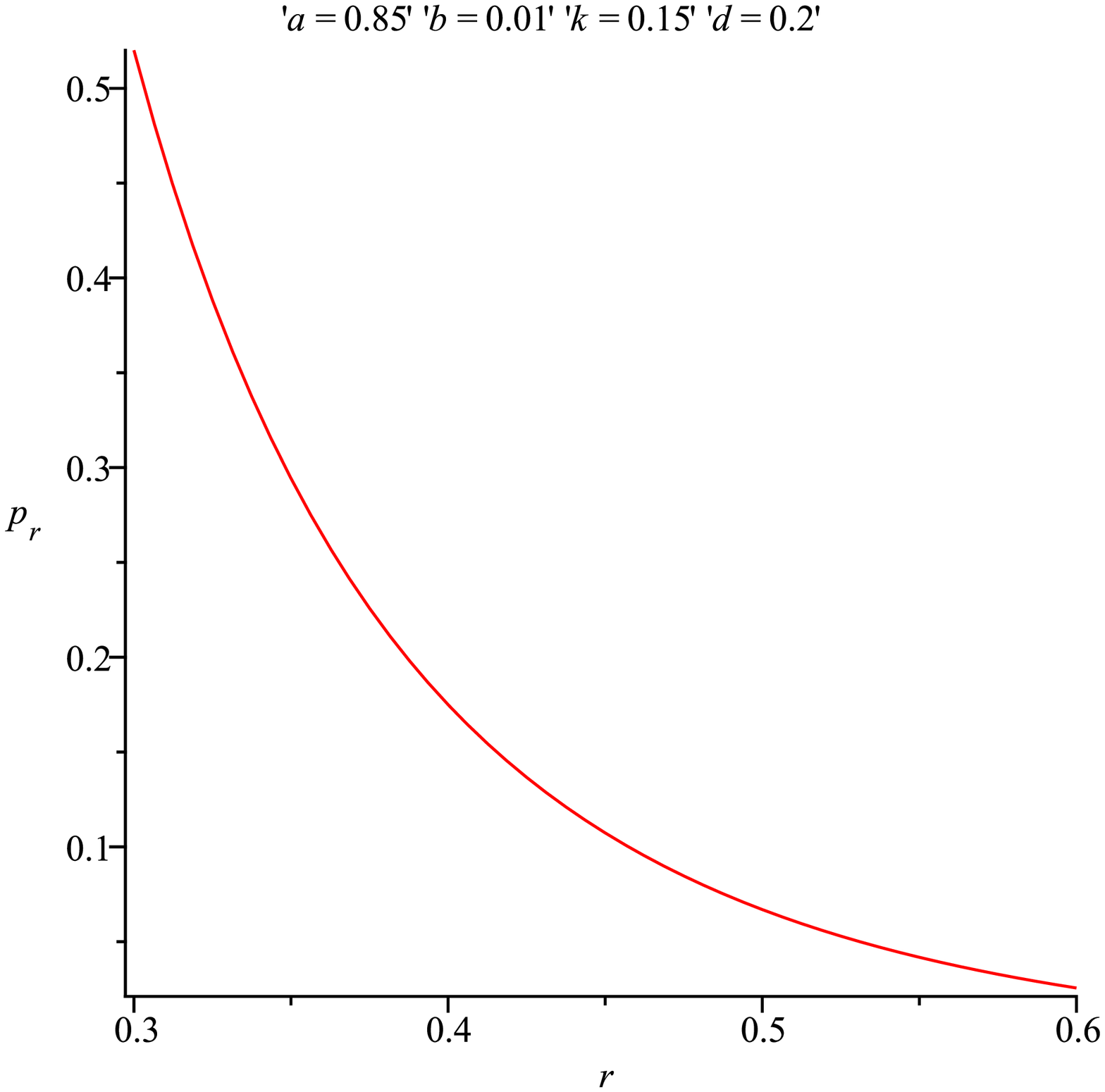}
\caption{Variation of radial pressure with respect to radial distance}
\end{figure}

\begin{figure}[htbp]
\includegraphics[scale=.3]{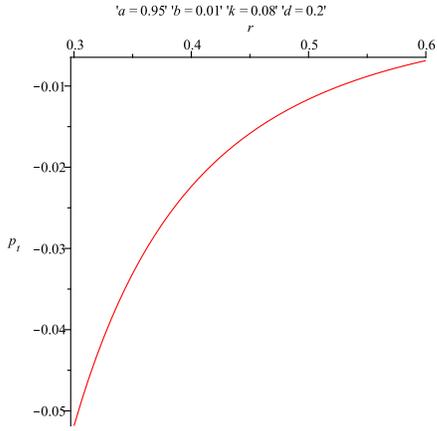}
\caption{Variation of transverse pressure with respect to radial distance}
\end{figure}

In Figs. 2 - 3 we have shown the behaviour of radial and
tangential pressures and whereas Fig. 1  represents the
$3$-dimensional presentation indicating the variation of radial
pressure with respect to the parameters $a,b$ for a fixed radius.
All the curves are in accordance to the physical feasibility
within the acceptable range.

Let us now consider a simplest form of barotropic equation of
state (EOS) as $p = \omega_i \rho$, where $\omega_i$ are the EOS
parameters along radial and transverse directions. Here at present
as such we are not considering different forms of $\omega$ with
it's possible space and time dependence as are available in the
literature
\cite{Chervon2000,Zhuravlev2001,Peebles2003,Usmani2008}. Thus, the
EOS parameters $\omega$ can straightly be written in the following
forms:
\begin{equation}
\omega_r= \frac{p_r}{\rho}= \frac{1}{(a+3br^2)} \left[2-3a-3br^2+
\frac{3d}{r} -2k \sqrt {g(r)}\right], \label{eq24}
\end{equation}

\begin{figure}[htbp]
\includegraphics[scale=.3]{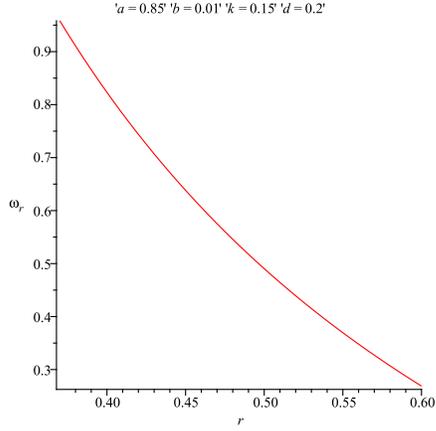}
\caption{Variation of radial EOS parameter with respect to radial distance}
\end{figure}

\begin{equation}
\omega_t= \frac{p_t}{\rho}= \frac{1}{(a+3br^2)} \left[1-a-3br^2+
k^2 - 2k \sqrt{g(r)}\right]. \label{eq25}
\end{equation}

In Figs. 4 and 5  we plot variations of radial and transverse EOS
parameter with respect to radial distance. In  which radial EOS
represents expected nature with a decreasing pattern. However,
transverse EOS parameter indicates that it lies between
$(-\frac{1}{3},~0)$. It is negative like exotic matter in nature,
however, later we will see that all energy conditions are
satisfied for real matter situation.

\begin{figure}[htbp]
\includegraphics[scale=.3]{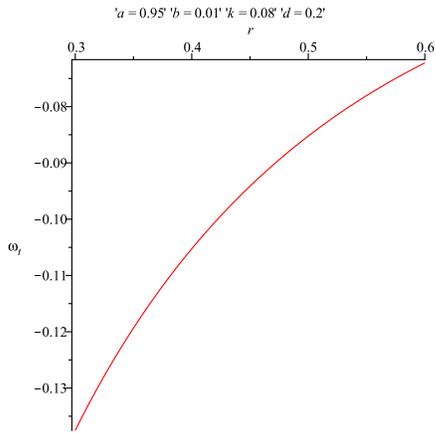}
\caption{Variation of transverse EOS parameter with respect to
radial distance}
\end{figure}

\subsection{Anisotropic behavior}

For the model under consideration the measure of anisotropy in pressures can be obtained,
from Eqs. (23) and (24), as follows:
\begin{equation}
\Delta\equiv(p_t-p_r)= \frac{1}{8\pi
r^2}\left[k^{2}+2a-\frac{3d}{r}-1\right]. \label{eq28}
\end{equation}

It can be seen that the `anisotropy' will be directed outward when
$p_t > p_r$ i.e. $\Delta
> 0 $, and inward when $p_t < p_r$ i.e. $\Delta < 0$. This feature is obvious from Fig. (5)
of our model that a repulsive `anisotropic' force ($\Delta < 0$)
allows the construction of a more massive stellar distribution
\cite{Hossein2012}.

\begin{figure}[htbp]
\includegraphics[scale=.3]{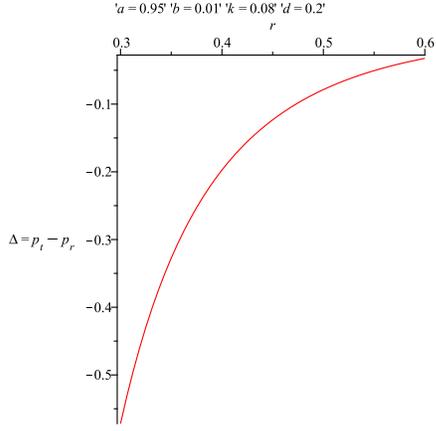}
\caption{Anisotropic behaviour at the stellar interior with respect to radial distance}
\end{figure}

\subsection{Energy conditions}

We now put and verify different energy conditions of the stellar configuration as follows:
\[\rho \geq 0, ~~~\rho+p_r \geq 0, ~~~\rho+p_t \geq 0,~~~\rho+p_r+2p_t \geq 0,\]
$\rho > |p_r|$ and $\rho > |p_t|$.

It is seen that Null Energy Condition (NEC), Weak Energy Condition
(WEC), Strong Energy Condition (SEC) and Dominant Energy Condition
(DEC) i.e. all the energy conditions for our particular choices of
the values of mass and radius are satisfied which can also be
observed from Fig. 7.

\begin{figure}[htbp]
\centering
\includegraphics[scale=.3]{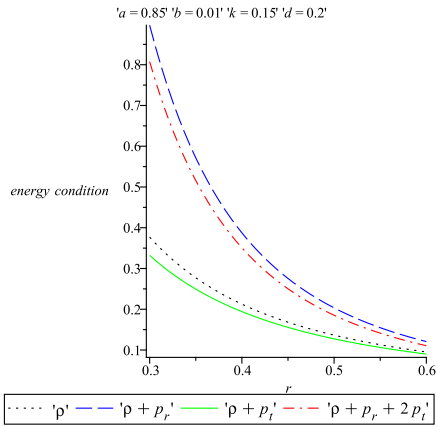}
\caption{Energy conditions as a function of $r$ are plotted for the specified
range}
\end{figure}

\subsection{TOV Equation}

To search equilibrium situation of this anisotropic star under
different forces, we write the generalized
Tolman-Oppenheimer-Volkoff (TOV) equation as  \cite{Varela2010}
\begin{equation}
-\frac{M_G(\rho+p_r)}{r^{2}}e^{\frac{\lambda-\nu}{2}}-\frac{dp_r}{dr}+\frac{2}{r}(p_t-p_r)=0,
\end{equation}
where $M_G=M_G(r)$ is the effective gravitational mass inside a
sphere of radius $r$ which can be derived from the
Tolmam-Whittaker formula as
\begin{equation}
M_G(r)=\frac{1}{2}r^{2}e^{\frac{\nu-\lambda}{2}}\nu'.
\end{equation}

The above equation explains the equilibrium conditions of the
fluid sphere due to the following hydrostatic, anisotropic and
gravitational forces:
\begin{equation}
F_h=-\frac{dp_r}{dr}=\frac{1}{8\pi}\left[\frac{4}{r^3}-\frac{6a}{r^3}
+\frac{9d}{r^4}-\frac{4k\sqrt{g(r)}}{r^3}-\frac{k(2br^{2}+\frac{d}{r})}
{r^3\sqrt{g(r)}}\right],\label{eqn28}
\end{equation}

\begin{equation}
F_a=\frac{2}{r}(p_t-p_r)=\frac{1}{8\pi}\left[-\frac{2}{r^3}+\frac{4a}{r^3}+\frac{2k^2}{r^3}
-\frac{6d}{r^4}\right],\label{eqn29}
\end{equation}

\begin{eqnarray}
F_g=-\frac{\nu'}{2}(\rho+p_r) \nonumber \\
=-\frac{1}{8\pi}\left[\frac{2}{g(r)r}
\{-\frac{12a}{r^2}+\frac{2}{r^2}+\frac{15d}{r^3}
-\frac{10k\sqrt{g(r)}}{r^2}+\frac{10ak\sqrt{g(r)}}{r^2}+\frac{18d^2}{r^4}
+\frac{6a^2}{r^2} \right. \nonumber\\ - \left.
\frac{15ad}{r^3}+6ab-6b-\frac{9bd}{r}+6bk\sqrt{g(r)}
-\frac{12dk\sqrt{g(r)}}{r^3}+\frac{4k^2 g(r)}{r^2}\}+\frac{a}{r^3}
-\frac{1}{r^3}-\frac{3d}{2r^4}+\frac{k\sqrt{g(r)}}{r^3}
\right],\label{eqn30}
\end{eqnarray}

\begin{figure}[htbp]
    \centering
        \includegraphics[scale=.3]{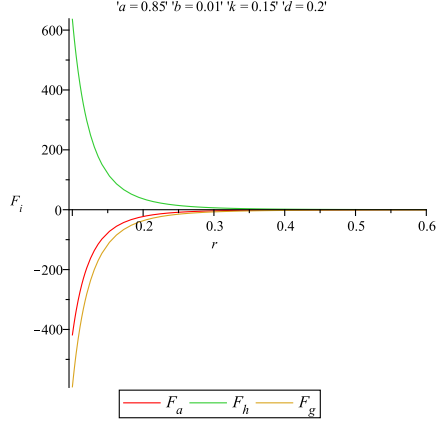}
       \caption{The dark energy star is in static equilibrium under gravitational
       $(F_g)$,hydrostatics $(F_h)$ and anisotropy $(F_a)$ forces.}
    \label{fig:3}
\end{figure}

The Eq. (28) can be rewritten in the form
\begin{equation}
F_g+F_h+F_a=0.
\end{equation}

The profiles of different forces $F_g$,~$F_h$,~$F_a$ are shown in
Fig. 8. The figure exposes that our model of  anisotropic star is
in static equilibrium under the gravitational $(F_g)$,
hydrostatics $(F_h)$ and anisotropic $(F_a)$ forces.

\subsection{Stability issue}

The velocity of sound should follow the condition $v_{s}^2 = dp/d\rho < 1$ for
a physically realistic model \cite{Herrera1992,Abreu2007}. We therefore calculate
the radial and transverse sound speed for our anisotropic model as follows \cite{Rahaman2012b}:
\begin{equation}
v_{rs}^2=\frac{dp_r}{d\rho}=\frac{1}{2a}\left[4-6a+\frac{9d}{r}
-4k\sqrt{g(r)}-\frac{k(2br^2+\frac{d}{r})}{\sqrt{g(r)}}\right],
\label{eq26}
\end{equation}

\begin{equation}
v_{ts}^2=\frac{dp_t}{d\rho}=\frac{1}{2a}\left[2-2a+2k^2 -4k
\sqrt{g(r)}- \frac{k(2br^2+\frac{d}{r})}{\sqrt{g(r)}}\right].
\label{eq27}
\end{equation}

\begin{figure}[htbp]
\includegraphics[scale=.3]{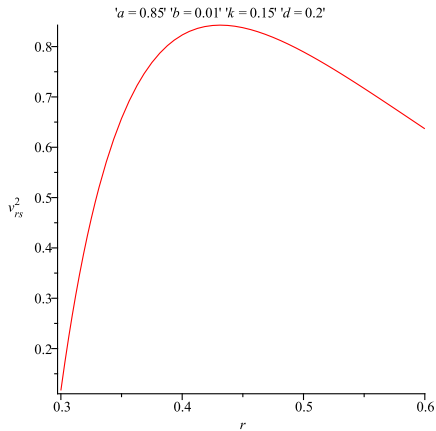}
\caption{Variation of radial sound velocity with respect to
radial distance.}
\end{figure}

\begin{figure}[htbp]
\includegraphics[scale=.3]{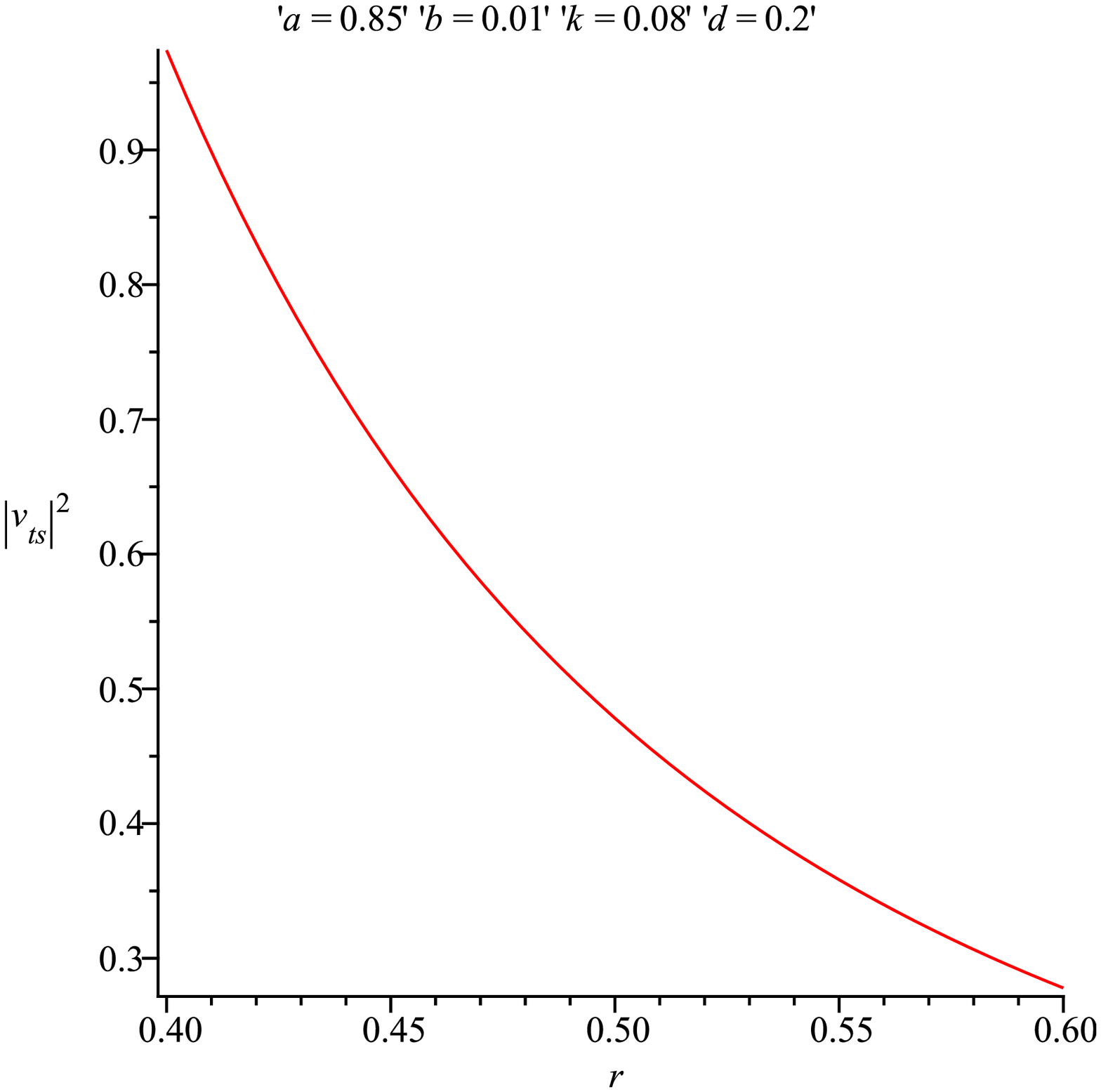}
\caption{Variation of transverse sound velocity with respect to
radial distance.}
\end{figure}

\begin{figure}[htbp]
\includegraphics[scale=.3]{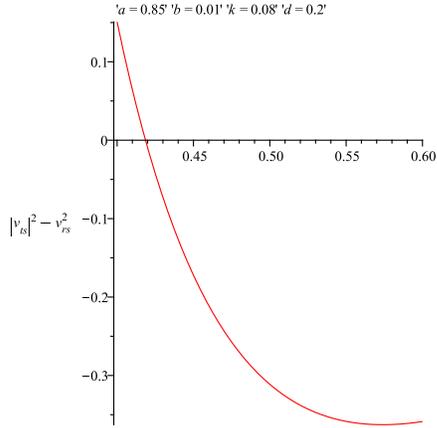}
\caption{Variation of $|v_{ts}^2| - v_{rs}^2 $ with respect to
radial distance.}
\end{figure}

Let us check whether the sound speeds lie between $0$ and $1$. For
this requirement we plot the sound speeds in Figs. 8 and 9. It is
observed that numerical values of these parameters satisfy the
inequalities $0 \leq v_{rs}^2,~|v_{ts}^2| \leq 1$ everywhere
within the stellar object. However, for radial sound velocity
there is a prominent change over after attaining certain higher
value which gradually goes down. Since the transverse sound
velocity is negative (due to negative transverse pressure), we use
numerical value. As sound speeds lie between $0$ and $1$, we
should have $|v_{ts}^2| - v_{rs}^2  \leq 1$ as evident from Fig.
10.

Now, one can go through a technique for stability check of local
anisotropic matter distribution as available in the literature
\cite{Herrera1992}. This technique is known as the cracking
concept of Herrera and states that the region for which radial
speed of sound is greater than the transverse speed is a
potentially stable region. Fig. 10 indicates that there is a
change of sign for the term $|v_{ts}^2| - v_{rs}^2$ within the
specific configuration and thus confirming that the model has a
transition from unstable to stable configuration. The present
stellar model gradually gets stability with the increase of the
radius. However, in terms of the maximum allowable compactness
(mass-radius ratio) for a fluid sphere as given by Buchdahl
\cite{Buchdahl1959}, the stability issue will be discussed later
on.

\subsection{Surface redshift}

One can calculate the effective gravitational mass in terms of the
energy density $\rho$ as
\begin{equation}
M_{eff}=4\pi\int_{0}^{R}\rho r^2 dr= \frac{1}{2}R\left[a+bR^2
\right]. \label{eq33}
\end{equation}

\begin{figure}[htbp]
\includegraphics[scale=.3]{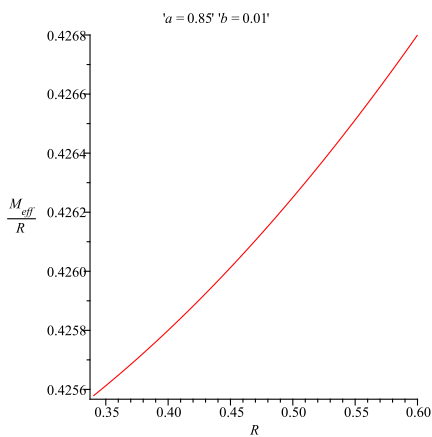}
\caption{Mass vs radius relation is shown in the plot for the specified
range}
\end{figure}

Therefore, the compactness of the star is given by
\begin{equation}
u= \frac{M_{eff}}{R}= \frac{1}{2}\left[a+bR^2\right].\label{eq35}
\end{equation}
The nature of variation of the above expression for compactness of
the star can be seen in the Fig. 9 which is gradually increasing.
We now define the surface redshift corresponding to the above
compactness as
\begin{equation}
 1+z_s= [1-2u]^{-1/2}=\left[1 - \left(a+bR^2\right)\right]^{-1/2}, \label{eq36}
\end{equation}
so that the surface redshift can be expressed as follows:
\begin{equation}
 z_s= \left[1-(a+bR^2)\right]^{-\frac{1}{2}} - 1. \label{eq36}
\end{equation}

\begin{figure}[htbp]
\includegraphics[scale=.3]{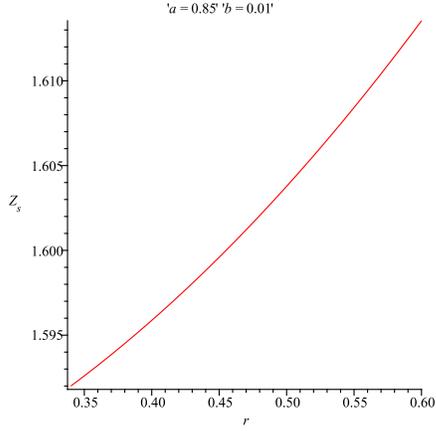}
\caption{Variation of redshift $z$ with radial coordinate $r$}
\end{figure}

We plot surface redshift in Fig. 9. Likewise compactness factor
$u$ this has also a behaviour of gradual increase. This feature is
also observed from the Table 1. The maximum surface redshift for
the present stellar configuration of radius $0.62$~km turns out to
be $z_s = 0.35280$ (see Table 1).

\begin{table}
\caption{Calculation of masses and hence compactness factors
for different values of constant $b$ with $a = 3/7$ and $R=0.62$~km
 (with the conversion $1~km = 1.475~M_{\odot}$)}
\vspace{0.50cm}
\begin{tabular}{|c|c|c|c|c|c|c|c|c|}
\hline
$b$ & 0.0 & 0.01 & 0.02 & 0.03 \\
\hline
$M_{eff}$(in $M_{\odot}$) & 0.19596 & 0.19952 & 0.20335 & 0.20717 \\
\hline
$u$ & 0.21428 & 0.21817 &  0.22236 & 0.22653 \\
\hline
$z_s$ & 0.32279 & 0.33257 & 0.34257 & 0.35280 \\
 \hline
\end{tabular}
\label{Table 1}
\end{table}

\section{SPECIAL PHYSICAL ANALYSIS OF THE MODEL BASED ON MATTER CONTAINED}

If we look at Eq. (35) as well as Eq. (36) then it will at once reveal that
there are lots of information inherently hidden inside these two equations.
Let us therefore examine the present model for the specified values of $a$ and $b$,
appearing in Eq. (35) and Eq. (36), as follows:

\subsection{The Misner prescription}
For the prescription of Misner \cite{Misner1964} i.e. $a = 3/7$
and $b = 0$, we get the effective gravitational mass as
$M_{eff}=(3/14)R$ which numerically comes out to be
$0.195~M_{\odot}$ so that the compactness factor becomes
$u=M_{eff}/R =0.21428$ as can be seen from Table 1. Here to
estimate mass we have taken radius of the star $R=0.62$~km by
solving $p_r=0$ at $r=R$ graphically from Fig. 11. This therefore
represents a relativistic Fermi gas for ultradense cores of
neutron stars as noticed in the Ref. \cite{Misner1964}.

\begin{figure}[htbp]
\includegraphics[scale=.3]{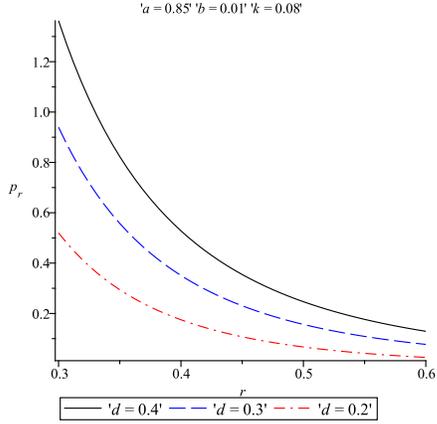}
\caption{Radius of the star where $ p_r=0$ increases with the increase of the parameter $d$}
\end{figure}

\subsection{The Dev-Gleiser prescription}
In this prescription \cite{Dev2004}, by substituting the values $a
= 3/7$ and $b \neq 0$ one can get the expression for the effective
gravitational mass as $M_{eff}= (1/2)R\left[(3/7)+bR^2 \right]$.
For this case we provide here a data sheet by choosing different
values for $b$ (for Case A, $b=0$ whereas for Case B, $b \geq 0$)
in Table 1. This corresponds to a relativistic Fermi gas with
core immersed in a constant density background \cite{Dev2004}.

\section{CONCLUDING REMARKS}

In the investigations we have observed some interesting points
which are as follows:

(1) Note that though transverse pressure is negative, however,
all energy conditions are satisfied. This indicates that matter
distribution of the star is real i.e. star is comprising of
normal matter.

(2) It is observed that the transverse sound velocity is negative
(due to negative transverse pressure) so that we use numerical
value. As sound speeds lie between $0$ and $1$, we should have $
|v_{ts}^2| - v_{rs}^2  \leq 1$ (Fig. 10).

(3) In connection to isotropic case and in the absence of the
cosmological constant it has been shown for the surface redshift
analysis that $z_s \leq
2$~\cite{Buchdahl1959,Straumann1984,Boehmer2006}. On the other
hand, B{\"o}hmer and Harko \cite{Boehmer2006} argued that for an
anisotropic star in the presence of a cosmological constant the
surface redshift must obey the general restriction $z_s \leq 5$,
which is consistent with the bound $z_s \leq 5.211$ as obtained by
Ivanov~\cite{Ivanov2002}. Therefore, for an anisotropic star
without cosmological constant the above value $z_s = 0.3$ seems to
be within an acceptable range (Fig. 12).

(4) The radius of the star depends on the parameter $d$, as
appears in Eq. (23), with an increasing mode (Fig. 13).

(5) It is well known that the anisotropic factor $\Delta=p_t -
p_r$ should vanish at the origin. However, in the present model
the energy density under consideration has an infinite central
density. According to Misner and Zapolsky \cite{Misner1964} one
may use it as an asymptotic solution with a cut-off density below
which the Eq. (27) will not valid.

(6) In the present model, the stability of the matter
distributions comprising of the anisotropic star has been
attained. We have come to conclude this by analyzing  the TOV
equation which describes the equilibrium condition for matter
distribution subject to the gravitational force ($F_g$),
hydrostatic force ($F_h$) and another force ($F_a$) due to
anisotropic pressure (Fig. 8).

(7) For an overall view of the present study and also to have a
physically viable model we put here the data which are available
on some compact stars, e.g. $Her~X-1$ and $SAX~J~1808.4-3658$ and
$Strange~Star-4U~1820-30$ (see Table 1 of the Ref.
\cite{Kalam2012}). The compactness factor $u$ for these stars are
$0.168$, $0.299$ and $0.332$ respectively. In comparison to these
data we can conclude that our model represents a star with
moderate compactness and also indicates that the star might be a
compact star of Strange/Quark type rather than Neutron star. This
is because, in general, superdense stars with mass-size ratio
exceeding $0.3$ are expected to be composed of strange matter
\cite{Tikekar2007}.

Now according to Buchdahl \cite{Buchdahl1959}, the maximum
allowable compactness for a fluid sphere is given by $\frac{2M}{R}
<\frac{8}{9}$. Our compactness values for the chosen model (see
Table 1) within this acceptable range and hence provides a stable
stellar configuration.

\section*{Acknowledgments}
FR and SR wish to thank the authorities of the Inter-University
Centre for Astronomy and Astrophysics, Pune, India for providing
the Visiting Associateship under which a part of this work was
carried out.

\end{document}